\documentclass[aip,apl,reprint]{revtex4-1} 

\usepackage{graphicx}
\usepackage{dcolumn}
\usepackage{bm}
\usepackage{amsmath, amssymb}
\usepackage{float}
\usepackage{mhchem}
\usepackage{eulervm}

\begin{document}
\bibliographystyle{unsrt}

\title{Control of 4-magnon-scattering in a magnonic waveguide by pure spin current}

\author{T. Hache}
\affiliation{Helmholtz-Zentrum Dresden--Rossendorf, Institute of Ion Beam Physics and Materials Research, 01328 Dresden, Germany}
\affiliation{Max Planck Institute for Solid State Research, 70569 Stuttgart, Germany}

\author{L. Koerber}
\affiliation{Helmholtz-Zentrum Dresden--Rossendorf, Institute of Ion Beam Physics and Materials Research, 01328 Dresden, Germany}
\affiliation{Technische Universit\"at Dresden, 01062 Dresden, Germany}

\author{T. Hula}
\affiliation{Helmholtz-Zentrum Dresden--Rossendorf, Institute of Ion Beam Physics and Materials Research, 01328 Dresden, Germany}
\affiliation{Institut f\"ur Physik, Technische Universit\"at Chemnitz, 09107 Chemnitz, Germany}

\author{K. Lenz}
\affiliation{Helmholtz-Zentrum Dresden--Rossendorf, Institute of Ion Beam Physics and Materials Research, 01328 Dresden, Germany}

\author{A. Kakay}
\affiliation{Helmholtz-Zentrum Dresden--Rossendorf, Institute of Ion Beam Physics and Materials Research, 01328 Dresden, Germany}

\author{O. Hellwig}
\affiliation{Helmholtz-Zentrum Dresden--Rossendorf, Institute of Ion Beam Physics and Materials Research, 01328 Dresden, Germany}
\affiliation{Institut f\"ur Physik, Technische Universit\"at Chemnitz, 09107 Chemnitz, Germany}

\author{J. Lindner}
\affiliation{Helmholtz-Zentrum Dresden--Rossendorf, Institute of Ion Beam Physics and Materials Research, 01328 Dresden, Germany}

\author{J. Fassbender}
\affiliation{Helmholtz-Zentrum Dresden--Rossendorf, Institute of Ion Beam Physics and Materials Research, 01328 Dresden, Germany}
\affiliation{Technische Universit\"at Dresden, 01062 Dresden, Germany}

\author{H. Schultheiss}
\affiliation{Helmholtz-Zentrum Dresden--Rossendorf, Institute of Ion Beam Physics and Materials Research, 01328 Dresden, Germany}
\affiliation{Technische Universit\"at Dresden, 01062 Dresden, Germany}

\date{\today}

\begin{abstract}

We use a pure spin current originating from the spin Hall effect to generate a spin-orbit torque (SOT) strongly reducing the effective damping in an adjacent ferromagnet. Due to additional microwave excitation, large spin-wave amplitudes are achieved exceeding the threshold for 4-magnon scattering, thus resulting in additional spin-wave signals at discrete frequencies. Two or more modes are generated below and above the directly pumped mode with equal frequency spacing. It is shown how this nonlinear process can be controlled in magnonic waveguides by the applied dc current and the microwave pumping power. The sudden onset of the nonlinear effect after exceeding the thresholds can be interpreted as spiking phenomenom which makes the effect potentially interesting for neuromorphic computing applications. Moreover, we investigated this effect under microwave frequency and external field variation. The appearance of the additional modes was investigated in the time-domain revealing a time delay between the directly excited and the simultaneously generated nonlinear modes. Furthermore, spatially resolved measurements show different spatial decay lengths of the directly pumped mode and nonlinear modes. 
 
\end{abstract}


\maketitle

\subsection*{Introduction}
Spintronic devices make use of the charge and spin of electrons. It is believed that the use of both degrees of freedom has a large potential to built novel communication and computing technologies \cite{arxiv.2301.06727,Dieny:2020aa}. 
Especially the development of neuromorphic computing is a current focus with the aim to exchange energy-demanding software with energy-efficient neuromorphic hardware\cite{Christensen_2022}. 
Spin waves (magnons) could be used to transmit information by frequency, amplitude, wavelength and phase between spintronic oscillators acting as neurons\cite{Grollier:2020aa}. In micro-structures and thin films they are excited by microwave magnetic fields, short laser pulses\cite{Kimel_2022} and spin-polarized\cite{Kiselev:2003aa} or pure spin currents\cite{Demidov:2012aa}. Highly coherent spin waves can be generated when microwave currents generate microwave magnetic fields. Laser pulses and spin currents usually excite a band of spin waves. Large spin-transfer (STT) or spin-orbit torques (SOT) can completely compensate the intrinsic damping torque of a ferromagnet\cite{ISI:000264630400001, SLONCZEWSKI1996L1} resulting in auto-oscillations of the magnetization which can show high coherency as well\cite{Zahedinejad:2020aa, arxiv.2301.03859}. 
\newline
Another approach of spin-wave generation is the use of nonlinear magnon-magnon interactions. Here, nonlinear magnons are created with different frequencies or wavelengths compared to the initially populated magnons. This is usually achieved in an efficient manner after exceeding a certain threshold amplitude of the initial magnons, for example, by large pumping powers via microwave magnetic fields\cite{PhysRevLett.122.097202}. \newline
In contrast, in this study nonlinear magnon generation is already achieved by pumping with moderate microwave magnetic fields in an ultra-thin magnonic waveguide due to strong reduction but without complete compensation of the damping by SOT. Therefore, the auto-oscillation regime is not yet reached, but the spin-wave amplitudes are increased for a wide band of frequencies. As demonstrated in previous studies, auto-oscillations generated above the threshold current show effects like injection-locking (synchronization) to the external microwave signal\cite{Synchronization_of_spin_Hall_nano-oscillators_to_external_microwave_signals, doi:10.1063/1.5082692, PhysRevApplied.13.054009, Wagner:2018aa}. Far below the threshold current, the damping is reduced and can increase the propagation lengths of spin waves\cite{doi:10.1063/1.4871519} significantly. Here, we focus on the regime slightly below the threshold current with strongly reduced damping. This regime was not yet intensively studied under additional microwave excitation. First, the pumping power needed to reach the nonlinear regime can be significantly reduced and, second, the process can be controlled by an additional SOT and, therefore, by an applied dc current. Above the thresholds of the nonlinear process, additional spin-wave frequencies are generated. The sudden appearance of these modes can be interpreted as a spiking process in separated frequency channels and, thus, might be interesting for neuromorphic computing. Moreover, it could be used to generate spin-wave frequency combs\cite{doi:10.1063/5.0090033} in ultra-thin magnetic microstructures, for frequency conversion in magnonics and has to be considered as an additional damping channel in magnonic waveguides with low or SOT-controlled damping. Furthermore, the use of SOT controlled nonlinear effects might be the key to efficiently couple spin waves as information carriers to optically readable spin-qubits\cite{https://doi.org/10.48550/arxiv.2208.09036, Simon:2022aa,Iacopo:uz,Lee-Wong:2020vc,Wang:2020vf} with the requirement of stable spin and charge properties close to the interface\cite{Neethirajan:2023aa}. 
\newline
First, an overview of the sample geometry and the calculated spin-wave dispersion is given. Second, it is shown how the spin-wave intensity evolves under SOT excitation (dc current) and microwave excitation, separately. As a next step, dc current and microwave excitations are combined to generate non-linear magnon modes via 4-magnon-scattering due to reduced damping conditions. It is shown that this effect is controllable by the microwave power or dc current magnitude. Furthermore, spatially-resolved, field-dependent and time-resolved measurements are conducted which give insight into the underlying processes of the magnon scattering process.

\subsection*{Device and Measurement geometry}

\begin{figure}[t]
\begin{center}
\scalebox{1}{\includegraphics[width=8.6 cm, clip]{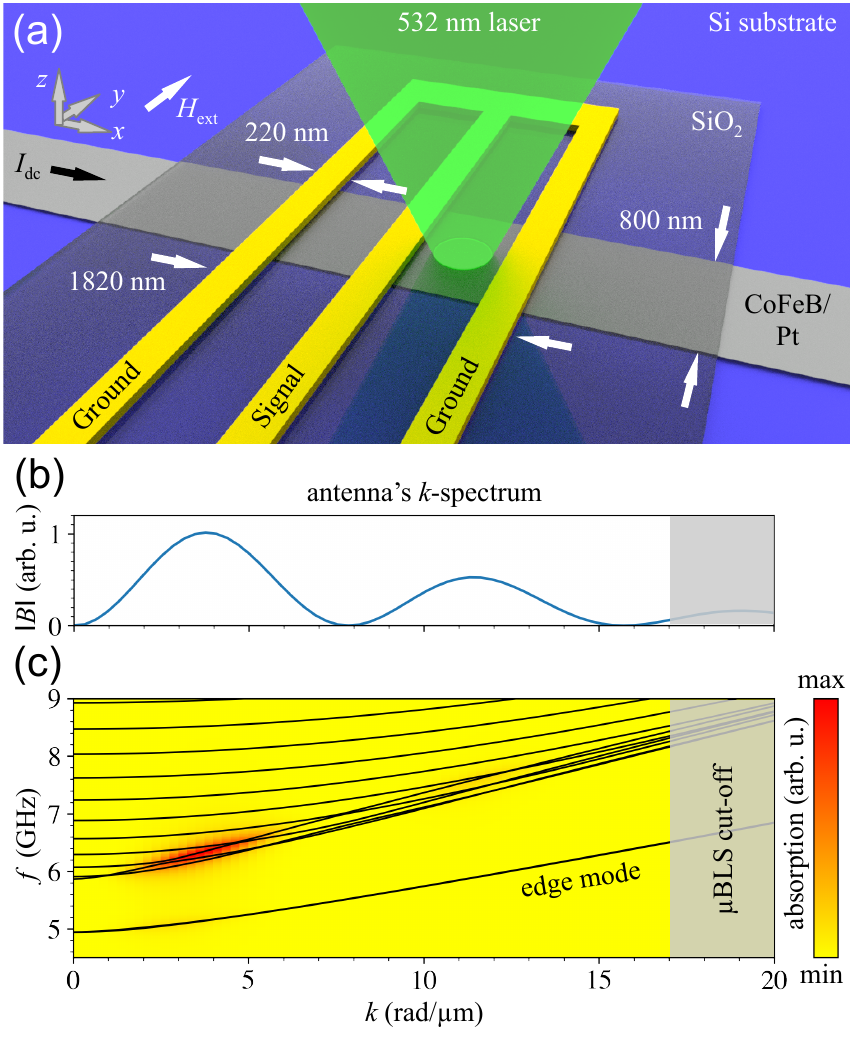}}
\caption{\label{fig1} (a) Sample and measurement geometry. The dc current is applied to the Pt/CoFeB stripe via gold contacts distanced by 9.5 µm. The \ce{SiO2} pad (40~nm thick) insulates the Pt/CoFeB stripe from the CPW antenna patterned on top.  (b) FFT of the field distribution of the antenna along \textit{x}-direction. (c) Spin-wave dispersion relation of the CoFeB stripe and absorption which marks the region of high pumping efficiency of the antenna field. The grey areas in c) and d) mark the cut-off \textit{k}-vector for the µBLS measurements limited by the numerical aperture of the used microscope lense. The external magnetic field is applied along the \textit{y}-axis and fixed at \textit{H}\textsubscript{ext}~=~43.6~mT for all measurements.}
\end{center}
\end{figure}

The sample under investigation is shown in Fig.~\ref{fig1}(a). A magnonic  waveguide consisting of a Ta(2)/Pt(7)/\ce{Co60Fe20B20}(5)/Ta(2) stack (numbers in nm) was fabricated on a high-resistive Si chip by means of electron beam lithography (EBL) and a subsequent deposition of the metallic layers via magnetron sputtering. An insulating \ce{SiO2} layer was fabricated by exposure of an HSQ resist in a second EBL step. The coplanar waveguide antenna (CPW) was patterned in a third EBL step with subsequent depositon of Cr(5)/Au(50) (numbers in nm) by thermal evaporation. \newline
An external magnetic field is applied perpendicular to the magnonic waveguide. A microwave current can be applied to the CPW by microwave probes. The generated Oersted field around the antenna excites the magnetization in the \ce{Co60Fe20B20} layer. Therefore, the excited spin waves propagate in Daemon-Eshbach geometry (spin-wave wave vector perpendicular to the equilibrium magnetization) along the magnonic waveguide except for the edges where the magnetization is tilted due to shape anisotropy. 
A dc current can be applied to the magnonic waveguide by additional dc contact pads (not shown) with a separation of 9.5 µm. Due to the amorphous structure of sputtered CoFeB films, the resistance is significantly higher than the one of the Pt layer. Therefore, most of the applied dc current flows in the Pt layer and generates a pure spin current in \textit{z}-direction via the spin Hall effect\cite{DyakonovPerrel,ISI:000082242600034,ISI:000324930200001} in the Pt layer. This pure spin current generates a spin-orbit torque\cite{9427163} in the adjacent \ce{Co60Fe20B20} layer and can increase or decrease the damping of the magnetization dynamics in dependence of the applied dc current polarity\cite{doi:10.1063/5.0008988}.   

Fig.~\ref{fig1}(b) shows the wave vector spectrum of the used CPW which was calculated by a Fourier transform of the antenna's Oersted field along the stripe in a distance of 40 nm (thickness of the \ce{SiO2}). Here, distinct maxima and minima are formed which have an impact on the excitable spin waves as discussed below. 
The spin-wave dispersion relation of the spin-wave modes propagating along the magnonic waveguide were calculated with a finite-element dynamic-matrix approach\cite{doi:10.1063/5.0054169} available in the TetraX micromagnetic-modeling package \cite{TetraX} and is shown in Fig.~\ref{fig1}(c). The following parameters were used: external magnetic field strength \textit{H}\textsubscript{ext}=~43.6~mT, saturation magnetization \textit{M}\textsubscript{S}~=~743~kA/m, exchange stiffness constant \textit{A}\textsubscript{ex}~=~0.87~pJ/m, Lande factor \textit{g}~=~2.14. The material parameters were determined by vibrating sample magnetometry (VSM) and ferromagnetic resonance measurements on a reference thin film. The exchange stiffness constant was determined from the perpendicular standing spin-wave mode. A Gilbert damping parameter ${\alpha}$~=~0.016 was measured without applied spin current on a reference thin film. The magnetic equilibrium configuration was obtained by initializing the magnetization parallel to the wavequide and subsequently minimizing the total magnetic energy including exchange and dipolar energy, as well as the Zeeman energy from the static external bias magnetic field. The various lines in Fig.~\ref{fig1}(c) belong to different transversally-standing modes which exhibit different numbers of nodal lines along the width direction of the waveguide. Additionally, the lowest-frequency curve describes the dispersion of the modes localized at the edges of the waveguide (edge modes). The color code of the graph encodes the absorption efficiency of the antenna's microwave field. The main absorption and, therefore, the most efficient excitation of spin waves is localized at wave vectors around 3.5~rad/µm (compare to Fig.~\ref{fig1}(b)) at a frequency of about 6.3~GHz. This is the branch of the fundamental mode (no nodal lines) along the stripe width. However, the confinement in space results in a broadening of the wave vector of all modes.\newline
In order to measure the spin waves in the waveguide, Brillouin light scattering microscopy was used (${\mu}$BLS)\cite{10.3389/fphy.2015.00035}. This technique is based on the inelastic scattering of light with magnons. 
The largest detectable wavevector is determined by the numerical aperture of the used microscope objective lense and reaches a value of about 17~rad/µm. This limit is sufficiently far above the range of the studied spin waves as shown as grey boxes in Figures~\ref{fig1}(b) and (c).\newline
\subsection*{Separate spin-orbit torque and microwave excitation}
\begin{figure}[t]
\begin{center}
\scalebox{1}{\includegraphics[width=8.6 cm, clip]{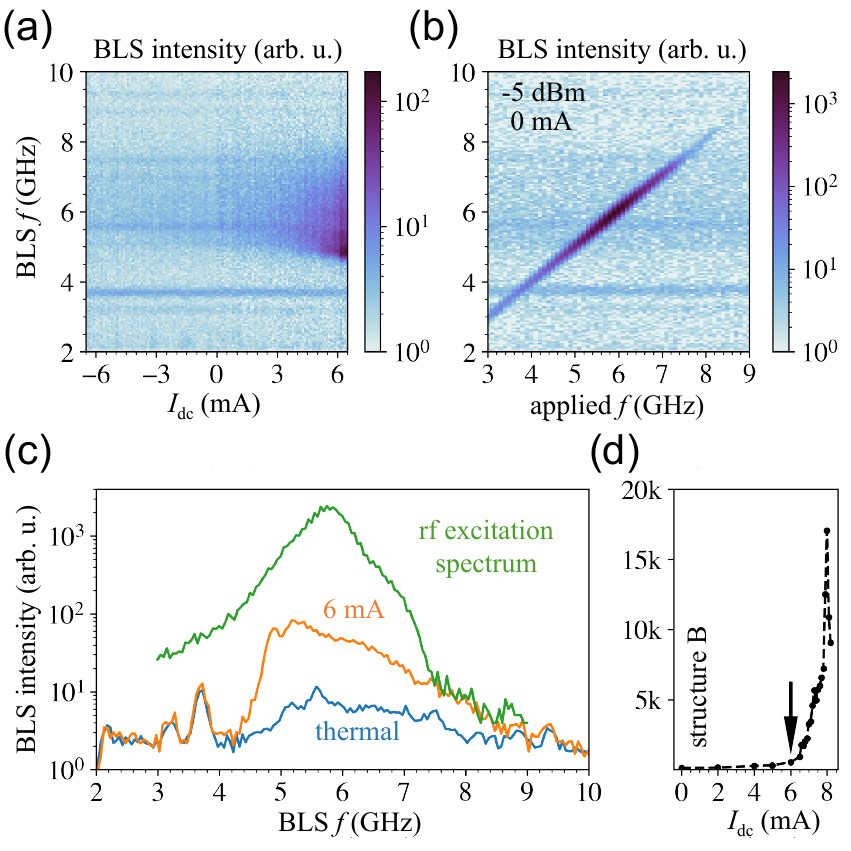}}
\caption{\label{fig2} (a) Spin-wave intensity in the Pt/CoFeB stripe upon dc current variation. An increase in spin-wave amplitude is achieved for positive dc currents only indicating the effect of the antidamping spin-orbit-torque. (b) Spin-wave intensity in the Pt/CoFeB stripe upon microwave excitation via the antenna without dc current. (c) Comparison of the excitation spectra (blue: thermal spin-wave spectrum, orange: spin-wave spectrum at 6~mA (from (a)), green: spin-wave intensity upon microwave excitation (from (b)) (d) Sweep of the dc current on an additional structure B which indicates the onset of auto-oscillations far above 6~mA used for the experiments of this manuscript.}
\end{center}
\end{figure}

As a first step, the spin-wave generation by dc current, microwave field and thermal excitation was measured separately to determine their impact on the magnonic system. Figure~\ref{fig2}(a) shows BLS spectra for various values of the applied dc current through the magnonic waveguide without any microwave excitation by the antenna. The measured spin-wave intensity increases only for positive dc current polarity which is expected by a pure spin current generated by spin Hall effect and resulting spin-orbit torque in this geometry. The excited spin-wave band is located between 4.5~GHz and 8~GHz. The upper limit is determined by the before mentioned largest detectable spin-wave wave vector in ${\mu}$BLS measurements. The lower limit is given by the lowest spin-wave mode in the waveguide.\newline 
%
Figure~\ref{fig2}(b) shows microwave excitation spectra in a frequency range  from 3~GHz to 9~GHz using the antenna without any applied dc current through the magnonic waveguide. As expected, the spin-wave intensity reaches a maximum within the measured spin-wave band of the waveguide. The spin-wave intensity was extracted and is plotted in Fig.~\ref{fig2}(c) together with a spectrum from Fig.~\ref{fig2}(a) at 6~mA and the spectrum of thermal spin waves (dc current and microwave at the antenna switched off). Please note that constant signals in the intensity maps correspond to side bands of the used laser only visible in the data due to logarithmic scaling. 
\newline
The largest spin-wave intensity during the microwave frequency sweep is reached at about 5.8~GHz. This is slightly lower in frequency than the point of largest absorption shown in Fig.~\ref{fig1}(c) at 6.3~GHz. This deviation is in range of the measurement uncertainty ($\pm$~5~\%) of the parameters used for the calculation. In addition, the transmission of microwave power might not be constant over the entire frequency range which can lead to a deviation as well. The spin-wave spectrum at 6~mA has a significantly different shape. Here, the highest spin-wave intensities are reached at the low frequency limit of the spin-wave band. This is in agreement with other experimental observations of SOT-excited spin waves and can be explained by the lower effective damping of the localized low frequency modes of the stripe\cite{ISI:000264630400001}. However, the SOT at these dc current values is not yet strong enough to compensate the damping completely and to achieve an auto-oscillation state with a large intensity of one spin-wave mode. The thermal spin-wave band (0~mA) has an about one order (two orders) of magnitude lower spin-wave intensity compared to the ones excited by dc current (microwave). In direct comparison to the spectrum at 6~mA, the spin-wave intensity at the lower frequency limit is not pronounced since the SOT is missing. \newline
For all experiments of this work it was important to set the dc current below the threshold of auto-oscillations. Figure~\ref{fig2}(d) shows the integrated spin-wave intensity measured on an additional device B. Here, the dc current was increased until breakdown of the device. It is concluded from this measurement that strong spin-wave amplitudes are reached at significantly higher dc currents above 7.4~mA which marks the onset of auto-oscillations. In all experiments shown in this manuscript, the dc current was set clearly below this limit (see black arrow) where the spin-wave intensity is significantly increased by SOT but without reaching the auto-oscillating state yet. In this regime, the damping is reduced and, therefore, allows to excite spin-wave modes to high amplitudes by microwave excitation via the CPW antenna as shown next. \newline
\subsection*{Nonlinear spin wave generation and control}
\begin{figure}[t]
\begin{center}
\scalebox{1}{\includegraphics[width=8.6 cm, clip]{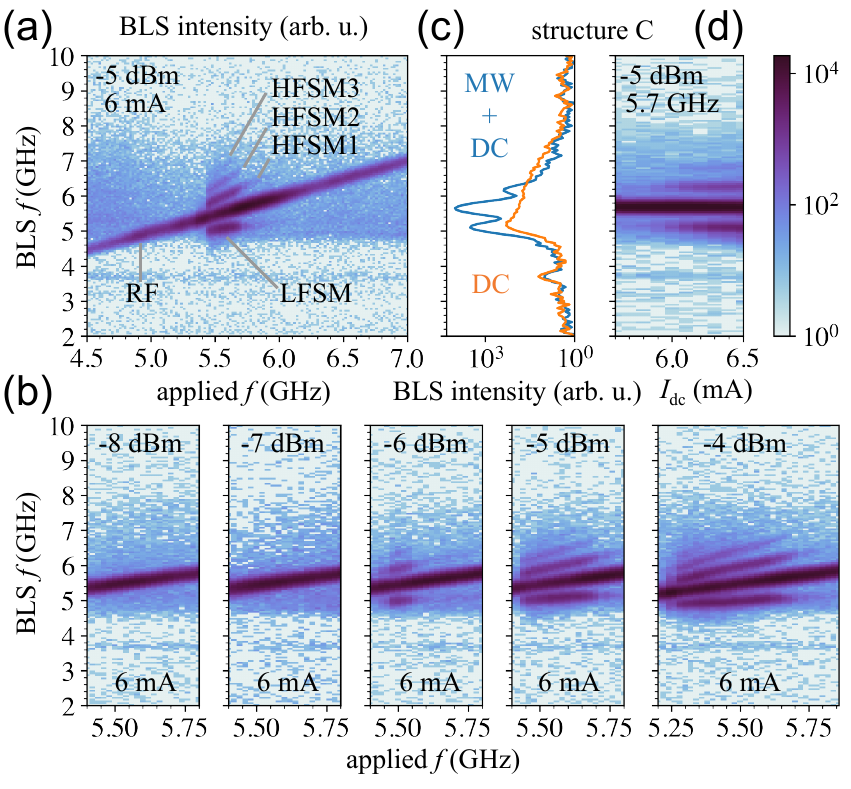}}
\caption{\label{fig3}(a) Sweep of the microwave frequency through the spin-wave band antidamped by the injected spin current. Additional modes appear in a range of pumping frequencies RF from 5.4~GHz to 5.8~GHz. One of the non-linear modes appears below (LFSM) and up to three non-linear modes are found above (HFSM1, HFSM2, HFSM3) the RF mode (To preserve the visual clarity the modes are labeled in (a) only). (b) Microwave power dependence. The additional modes appear only at a threshold microwave power above -7~dBm. The frequency range in which side modes appear increases for higher pumping powers. The antennas break at powers above -4~dBm. (c) Comparison of dc current excited spin-wave spectrum at 6.5~mA (orange) with the spectrum upon additional microwave excitation (blue) at 5.66~GHz, -5~dBm (structure C). The lowest split mode frequency seems to coincide with the intensity maximum of the spin-wave band. (d) Sweep of the dc current. The additional modes appear above a dc current threshold of about 5.8~mA (structure C). All intensity maps are normalized to the same color scale.}
\end{center}
\end{figure}

The data shown in Fig.~\ref{fig3} was acquired for a dc current of 6~mA through the magnonic waveguide and additional excitation of the spin-wave system via the CPW antenna. Figure~\ref{fig3}(a) shows spin wave spectra in an excitation-frequency range from 4.5~GHz to 9~GHz at constant microwave power (-5~dBm) and dc current. Compared to the measurement without dc current (Fig.~\ref{fig2}(b)), here, additional side modes appear for microwave frequencies from 5.4~GHz to 5.8~GHz. The mode RF is the directly excited mode at the applied microwave frequency. One mode is found below the RF mode (low frequency split mode: LFSM) and up to three modes are detected above the direct excitation (higher frequency split modes: HFSM1, HFSM2, HFSM3). To preserve the visual clarity the modes are labeled in Fig.~\ref{fig3}(a) only. The separation of the peaks ranges from 400~MHz to 700~MHz as function of the applied microwave frequency. The range of microwave frequencies showing this nonlinear effect can be controlled by the microwave power applied to the CPW antenna. As shown in Fig.~\ref{fig3}(b), below -6~dBm the effect is not observed but manifests clearly at higher powers. At -6~dBm the effect is maintained about 100~MHz and at -4~dBm over a range of about 500~MHz. At even higher microwave powers, the CPW antennas broke due to Joule heating. \newline
%
%
Figure~\ref{fig3}(c) shows the comparison of the spin-wave spectrum excited with dc current at 6.5~mA solely (orange) and the spectrum with additional microwave excitation of 5.66~GHz and a power of -5~dBm at the same dc current on an additional structure C. First, it reveals that the frequency of the LFSM is about the frequency of the intensity maximum of the spin-wave spectrum for pure dc current excitation. Second, the intensity at this frequency is increased by a factor of 16 due to the nonlinear process. This can be explained by a transfer of power from the directly pumped mode to the LFSM.
Besides the microwave power this nonlinear effect can be controlled by the dc current applied to the magnonic waveguide as shown in Fig.~\ref{fig3}(d). Here, the microwave power and frequency (-5~dBm, 5.7~GHz) was kept constant and the dc current was swept from 5.6~mA to 6.5~mA (measured on structure C). It is clearly visible that the additional side peaks appear after a certain threshold current of about 5.8~mA. Therefore, this nonlinear effect is activated after exceeding a certain level of antidamping SOT simply controlled by the dc current. \newline
\subsection*{Spatially-resolved measurements}
In order to gain further insight into the mode properties, spatially resolved ${\mu}$BLS measurements were performed across and along the magnonic waveguide. Here, the spin-wave intensity distribution was first measured during pure dc current (6.5~mA) excitation and afterwards with additional microwave excitation (5.66~GHz, -5~dBm) on structure C. \newline
\begin{figure}[t]
\begin{center}
\scalebox{1}{\includegraphics[width=8.6 cm, clip]{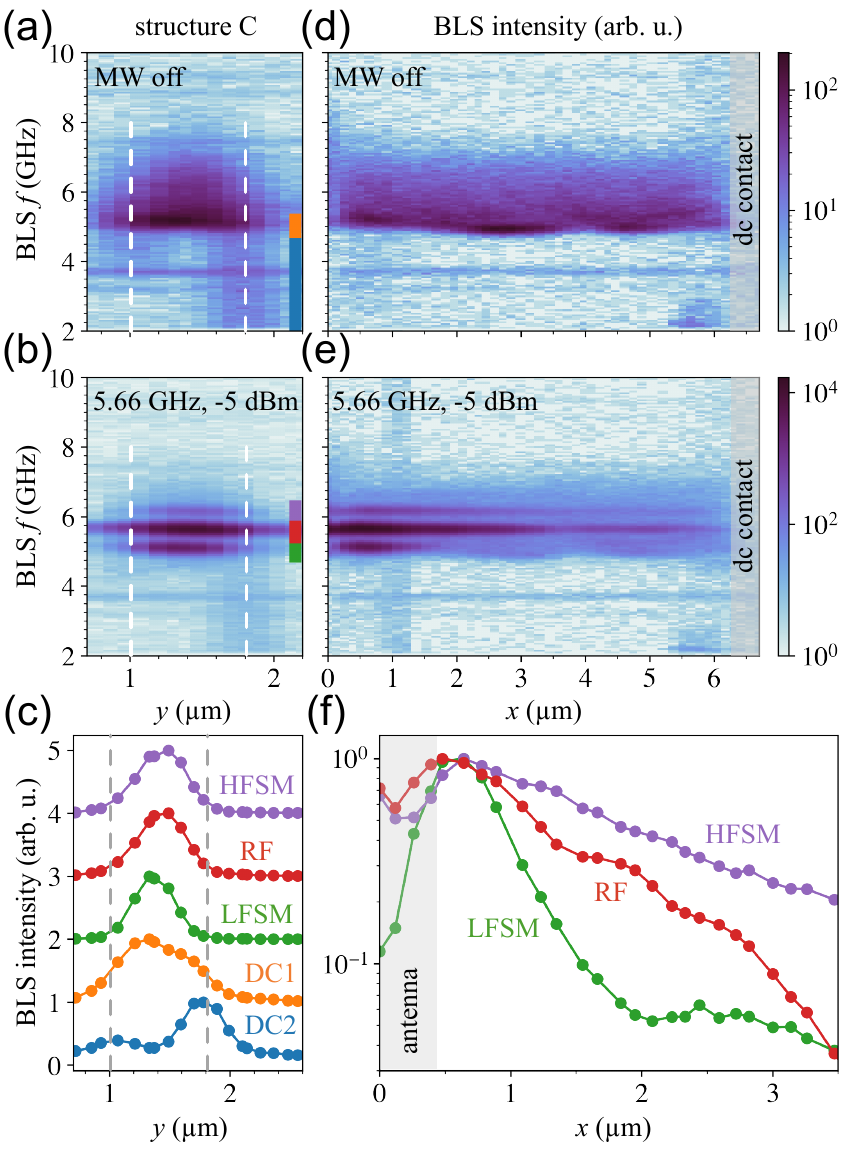}}
\caption{\label{fig4} Spatial distribution of the spin-wave intensity in the CoFeB stripe at 6.5 mA (a) Spin-wave intensity across the stripe during pure dc current excitation (b) Spin-wave intensity across the stripe upon dc current and microwave excitation (c) Comparison of integrated spin-wave intensity distributions. DC1: intensity above 4.8~GHz, DC2: intensity below 4.8~GHz in (a). RF: intensity of directly excited mode, LFSM: intensity of lower frequency split mode, HFSM: intensity of higher frequency split mode in (b). Integration ranges are given as colored segments in (a) and (b). The dashed lines indicate the edges of the CoFeB/Pt stripe. (d) Spin-wave intensity along the stripe upon dc current excitation. Local intensity (frequency) variations appear due to differences in local current densities, heat and Oersted fields. (e) Spin-wave intensity along the stripe upon dc current and microwave excitation. (f) Integrated spin-wave intensities of the modes in (e). The LFSM appears to stay localized near the antenna. The directly excited mode (RF) and the HFSM have a further extent.}
\end{center}
\end{figure}
Figure ~\ref{fig4}(a) shows the spin-wave intensity across the stripe during pure dc current excitation. The dashed lines mark the edges of the stripe. Clearly, the main intensity is located above 4.8~GHz (DC1) and is centered within the stripe. Below, the intensity is weaker and more localized at the edges (DC2). To clarify the distribution of the spin-wave intensities the data was integrated for both frequency ranges and is plotted in Fig.~\ref{fig4}(c). Here, the different extracted intensity distributions were normalized and shifted vertical equidistantly. The occurrence of spin-wave modes in the center and at the edges is in agreement to other investigations of modes in SHNO nanowire geometries\cite{PhysRevB.102.054422}. Due to fabrication defects like rough wire edges, the spin-wave intensity is smeared over a wide frequency range and is asymmetric. The DC1 mode in contrast is more localized in the wire center. Due to the limitation of spatial resolution caused by the width of the used laser spot of about 320 nm, the spin-wave mode profile can not be further resolved.\newline
Figure~\ref{fig4}(b) shows the spin-wave intensity at the same position during additional microwave excitation to reach higher spin-wave intensities and to enable the nonlinear effect. Besides the directly pumped mode RF, the non-linear low frequency split mode (LFSM) and the non-linear high frequency split mode (HFSM) are clearly visible. The mode profiles were extracted by integration of the spin-wave intensities and are plotted in Fig.~\ref{fig4}(c). These mode profiles clearly indicate the localization of the modes in the middle of the magnonic waveguide. This is a first hint that the modes at the edges of the stripe do not take part in the non-linear process of magnon generation.
\newline
Figure~\ref{fig4}(d) shows the spin-wave intensity along the magnonic waveguide during pure dc current excitation. The detected spin-wave band between 4.5~GHz and 8~GHz varies slightly in frequency and intensity along the stripe. We assume that small variations of the current densities and magnetic properties are induced by the fabrication process. Additionally, it is expected that the temperature varies slightly along the stripe causing frequency and intensity variations.\newline
\begin{figure*}[t]
\begin{center}
\scalebox{1}{\includegraphics[width=17.5 cm, clip]{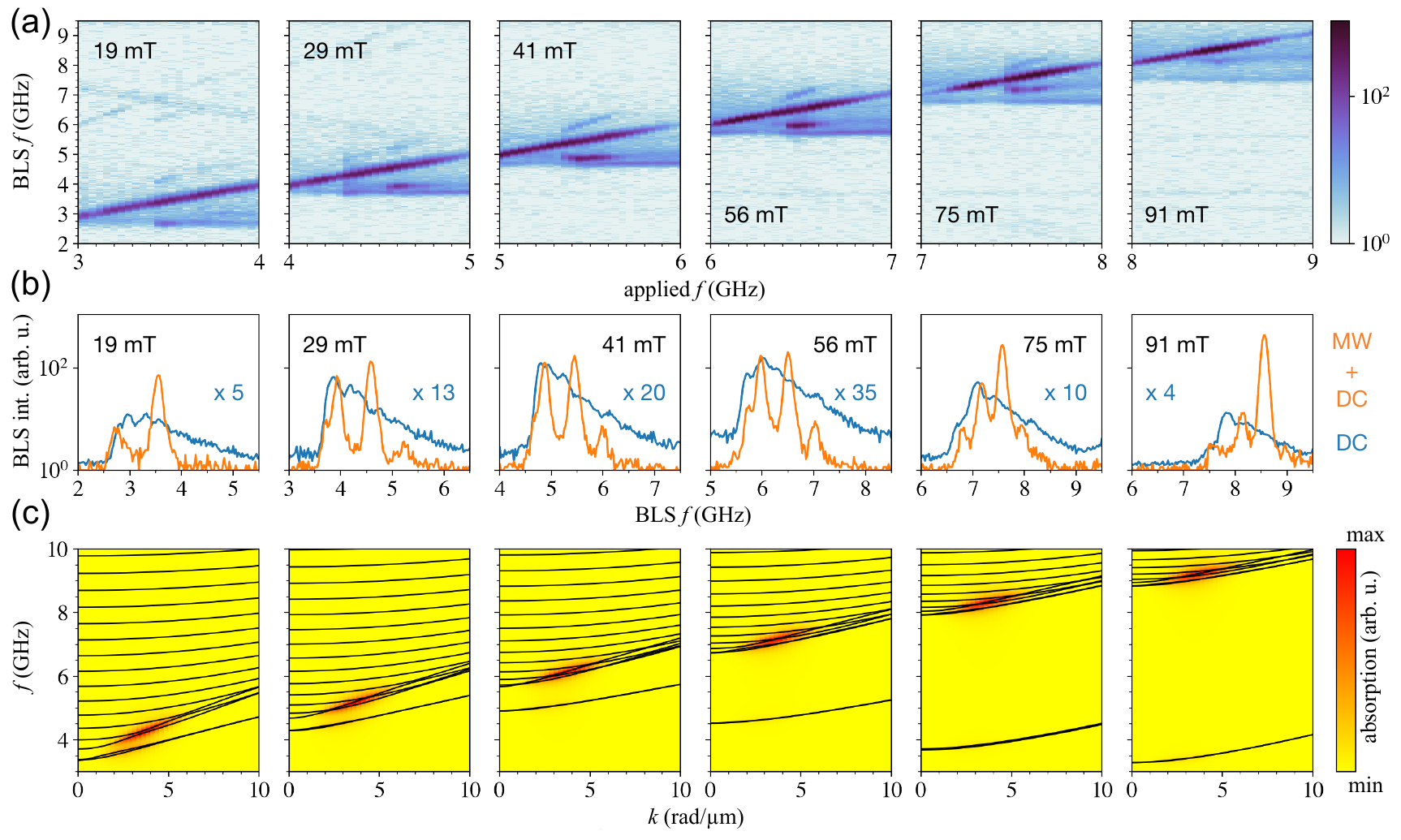}}
\caption{\label{fig7} Field-dependent measurements on structure D. (a) Microwave frequency sweeps at given external magnetic fields at 6.6~mA and -5~dBm microwave power. (b) Spin-wave spectra at the microwave frequencies with the highest intensity of the spin-waves in comparison to the spin-wave spectra without microwave excitation. The spin-wave spectra without microwave excitation were multiplied by a factor given in the graphs. The directly excited mode RF and the nonlinear modes are distributed in the same frequency range as the modes excited by the dc current only. The nonlinear modes are spaced equally around the corresponding directly excited RF mode. (c) Numerically- calculated spin-wave dispersion relations and absorptions of microwave power for the external field values shown above. The quasi-uniform mode is excited strongest by the microwave antenna and shows the same frequency dependence as the excited and nonlinear modes in (a). The frequencies of the edge modes in contrast show a different field dependence.}
\end{center}
\end{figure*}
Figure~\ref{fig4}(e) shows the spin-wave intensity along the stripe during dc current and microwave excitation. The largest spin-wave intensity is reached in close vicinity to the antenna located at about \textit{x}~=~0~µm and reaches significantly higher values than for the pure dc current excitation shown before. The spin-wave intensities along the stripe were integrated and normalized for all three modes RF, LFSM and HFSM and are compared in Figure~\ref{fig4}(f). The LFSM shows a stronger decay compared to both other modes. Reasons could be a smaller group velocity or the localization due to the formation of a local spin-wave potential well at the low frequency limit of the spin-wave band. Interestingly, the HFSM shows a weaker decay then the directly excited RF mode. A reason could be a stronger damping of the RF mode due to the flow of power to both split modes which takes place not only at the antenna position but over some distance from it. Please note that the increase of spin-wave intensity of the LFSM mode at about \textit{x}~=~2.5~µm is caused by the pure dc excitation which overlaps with the same frequency channels (compare local spin-wave intensity maximum in Fig.~\ref{fig4}(d) and (e).            
\subsection*{Magnetic field dependence}
In addition to the spatially-resolved measurements the non-linear generation of magnons was studied at different external magnetic fields to get a deeper inside in the involved magnon modes. The measurements were conducted on a separate structure D. 
The dc current was kept constant for all measurements at 6.6~mA. The microwave power was fixed at -5~dBm. The microwave frequency range had to be adjusted for all field values since the spin-wave dispersion shifts to higher frequencies for larger external fields. This shift is clearly visible in of Fig. ~\ref{fig7}(a). The spin-wave band gap at the lower frequencies (no spin-wave intensity) increases with external field from 2.6~GHz at 19~mT to 7.5~GHz at 91~mT. Nonlinear split modes were detected symmetrically separated around the directly excited mode for almost all fields. Only for the lowest and the largest external field values the intensity of the effect is reduced significantly. This can be understood in a qualitative fashion by having a look at the antidamping and damping effects influencing the spin-wave intensity of the directly pumped mode. Small external fields might be insufficient to align the magnetization perpendicular (parallel) to the applied dc current (injected spin current polarization) which reduces the antidamping and, therefore, decreases the spin-wave intensity weakening the nonlinear effect. At high external fields the magnetization is mainly aligned perpendicular to the stripe and the antidamping SOT is maximized. However, with increasing field the Gilbert damping torque is enhanced and lower spin-wave intensities are reached. This is in agreement with the magnetization's equilibrium alignments determined from the calculations. The width of the saturated region where the normalized perpendicular magnetization component is larger than 0.9999 was extracted for all fields from 19 mT to 91 mT: 352, 468, 544, 600, 664, 788 (in nm). Thereby, this component reaches the following minimum values at the waveguide edges, respectively: 0.4510, 0.5688, 0.6883, 0.8182, 0.9649, 0.9986.
Only for intermediate external fields the antidamping effect of the SOT is maximized and the Gilbert damping torque is still in a range which allows for high spin-wave intensities needed for nonlinear effects. \newline
In Fig. ~\ref{fig7}(b) the spectra with the highest spin-wave intensity of the nonlinear modes is compared to the spin-wave spectra without microwave but dc current excitation. This comparison demonstrates that the nonlinear modes coincidence mostly with the same frequency range of spin-waves excited by the dc current solely. Moreover, the nonlinear modes are separated symmetrically around the directly excited modes for all field values indicating underlying 4-magnon-scattering processes. \newline
Furthermore, it can be seen that the LFSM does not always coincident with the lowest frequency mode of the spin-wave spectra excited by dc current. Interestingly, two nonlinear modes below the frequency of the microwave-pumped mode  appear for 75~mT and 91~mT. Possibly, the first LFSM below the directly pumped mode reaches spin-wave intensities which result in a second order 4-magnon-scattering process generating a magnon at a lower frequency and at the frequency of the directly pumped mode RF. 4-magnon-scattering as possible explanation for the shown effect will be discussed at a later point in more detail. 
\newline
Figure ~\ref{fig7}(c) shows the spin-wave dispersion relations calculated with TetraX for the external fields used in the experiments. For all values the quasi-uniform mode of the stripe is excited most efficiently. The frequency of the quasi-uniform modes increases for larger external field values in a similar way as the frequencies of the directly excited modes and the nonlinear modes in the experiment (Fig.~\ref{fig7}(a)). The lowest frequency branches in the dispersions are excitations at the edges of the stripe where the magnetization is not aligned with the external magnetic field as mentioned before. Up to 41~mT the frequency of the edge modes increases and decreases from 56~mT to 91~mT. Thus, the edge modes clearly show a different frequency behavior than the excited and nonlinear modes studied in the experiment. This behavior is expected for magnetic stripes with perpendicular magnetic field before reaching the saturation field. Due to the frequency reduction of the edge modes, it is concluded that they don't take part in the nonlinear magnon scattering process. This is a further confirmation in addition to the spatially-resolved measurements shown before which showed the localization of the direct excited RF mode and the non-linear modes LFSM and HFSM in the center of the waveguide.

\subsection*{Time-resolved measurements}
%
\begin{figure}[b]
\begin{center}
\scalebox{1}{\includegraphics[width=8.6 cm, clip]{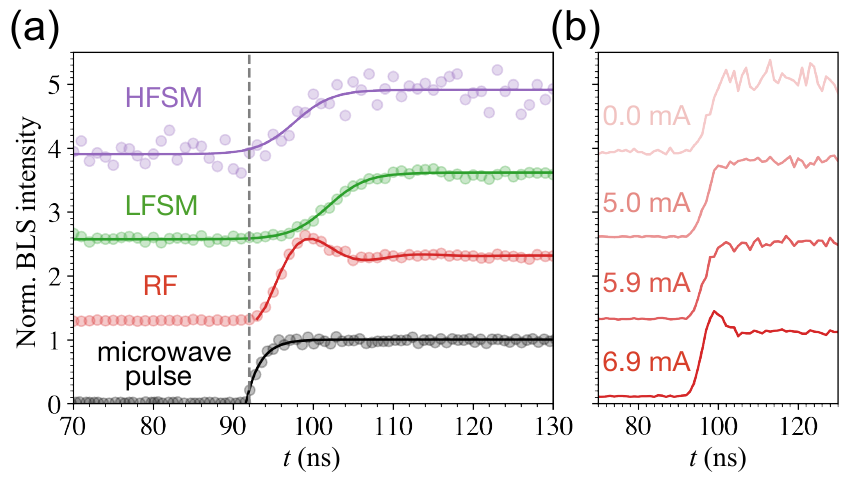}}
\caption{\label{fig6} Time-resolved µBLS intensity on structure E. (a) Time-dependence of the directly excited (RF) and nonlinear modes (LFSM, HFSM) at 6.9~mA, 5.62~GHz and -4~dBm (application of the microwave pulse at about 92~ns). LFSM and HFSM were fitted by a sigmoidal function, RF was fitted by an overshoot oscillation equation, the microwave pulse was fitted by an exponential function. Rise times: microwave pulse about 4 ns, RF mode: 10 ns, LFSM: 10 ns, HFSM: low SNR (b) Integrated spin-wave intensity of RF mode at different dc currents. An overshoot is only detected for high dc values when the SOT antidamps the system significantly.}
\end{center}
\end{figure}
In order to have a closer look to the process which generates the nonlinear modes, the microwave excitation was pulsed and the spin-wave spectra were measured time-resolved in a stroboscopic fashion on an additional device E. The microwave was switched on for 1~µs and off for 1~µs with a repetition rate of 0.5~MHz. 
The spin-wave intensities of the directly excited RF mode and the nonlinear LFSM and HFSM were normalized to the steady state intensity and are plotted in Fig.~\ref{fig6}(a) as red, green and violet dots, respectively. The traces are shifted equally along the \textit{y}-axis of the plot. The gray dashed vertical line marks the onset of the RF mode which is excited by the microwave voltage pulse (envelope, rectified, black dots). After a strong intensity increase of the RF mode and an overshoot the intensity reaches the steady state after about 13 ns. The shape of the microwave pulse does not show such an overshoot and is, therefore, not the reason. In contrast, the LFSM  intensity starts to strongly increase at a later time when the direct RF mode shows the overshoot at about 100 ns. Therefore, it can be concluded that the direct excitation RF has to exceed a critical power threshold in the first place before the nonlinear effect takes place efficiently. Due to the flow of power to the secondary modes (LFSM, HFSM), the intensity of the RF mode decreases until a steady state is reached. The microwave pulse, the RF and the LFSM mode were fitted by an exponential, overshoot and sigmoidal function, respectively, to extract the rise times. The microwave pulse has the shortest rise time of about 4 ns. The RF and LFSM mode have both a rise time of about 10 ns. Figure~\ref{fig6}(b) shows that the overshoot of the RF mode disappears by reducing the dc current causing a lower SOT and, therefore, larger effective damping which results in lower spin-wave intensities of the directly pumped RF mode. As result, no nonlinear modes appear and no overshoot was measured for smaller dc currents. \newline
\subsection*{Discussion}
\begin{figure}[b]
\begin{center}
\scalebox{1}{\includegraphics[width=8.6 cm, clip]{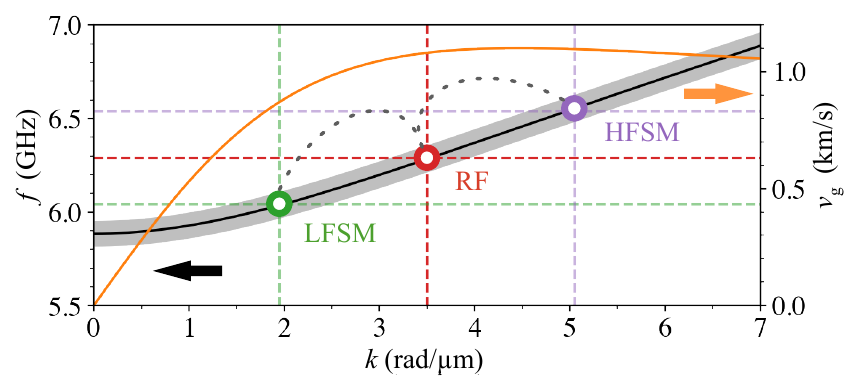}}
\caption{\label{fig5} Spin-wave branch (black) with fundamental quantization across the magnonic waveguide which is strongly excited at 3.5~rad/µm by the CPW antenna. The 4-magnon-scattering process takes place at a linear part of the dispersion since energy (\textit{f}) and momentum (\textit{k}) have to be conserved. The split modes LFSM and HFSM are, therefore, separated equidistantly around the directly excited mode RF in \textit{f} and \textit{k}. The group velocity (orange) of the spin-waves indicates significant lower values for the LFSM which fits to the findings of the spatially resolved measurements. The RF and HFSM mode have a larger group velocity.}
\end{center}
\end{figure}

The experimental observations indicate 4-magnon scattering process generating the nonlinear spin wave modes. A schematic of this process is shown in Fig.~\ref{fig5}. The spin-wave branch with the fundamental quantization over the magnonic waveguide's width is shown in black. As calculated in Fig.~\ref{fig1}(c) and Fig.~\ref{fig7}(c), this branch is excited strongest by the CPW antenna at a wave vector of 3.5~rad/µm. Therefore, it is expected that only there the spin-wave intensity can be increased to a level which makes nonlinear processes probable. Two of the directly excited RF magnons interact and create a new magnon LFSM with a lower frequency and a new magnon HFSM with a higher frequency. It is assumed that both nonlinear modes are also part of the shown spin wave branch because the spatially-resolved measurements shown in Fig.~\ref{fig4}(b) revealed the same symmetry and a location in the center of the magnonic waveguide. Due to the conservation of energy, both created magnons must be separated equidistantly around the RF mode in frequency space as it was found in the experiments. Additionally, the conservation of momentum demands for an equidistant separation in wave vector space. 
Based on this demands, the 4-magnon scattering process can only appear at a part of the spin-wave branch with linear slope. Indeed, the dispersion relation shows a quite linear monotonic increase around the pumped RF mode. The grey area visualizes qualitatively the broadening due to confinement and inperfections of the waveguide properties. As result, this process can happen not only by hitting the perfect conditions but also slightly around them. This explains qualitatively the findings of  the microwave power dependent measurements shown in Fig.~\ref{fig3}(b). At low power the scattering process appears only by hitting the perfect conditions (-6~dBm). But at larger powers a certain mismatch in frequency still allows for the scattering process due to the line width of these modes and their acceptable overlap.
The derivative of the spin-wave branch results in the group velocity shown in orange. Due to the flattening of the spin-wave dispersion at low \textit{k}-values, the group velocity of these spin waves is reduced compared to the overs. Therefore, it would be expected that the LFSM mode shows a reduced propagation in comparison to the other modes RF and HFSM. Indeed, this was found by measuring the decay in Fig.~\ref{fig4}(f). The LFSM mode shows a much faster drop of spin-wave intensity than the RF and HFSM mode. Additionally, the calculation revealed a slightly larger group velocity for the HFSM compared to the RF mode. This slight difference was also seen in the experiments. 
\subsection*{Conclusion}
4-magnon-scattering can be controlled in magnonic waveguides by a pure spin current generated via spin Hall effect. On the one hand the strength of the injected spin current and resulting SOT magnitudes do not enable the formation of magnetic auto-oscillations. On the other hand it decreases the damping of the magnonic system significantly. This results in increased spin-wave intensities in a wide frequency range. The additional pumping with microwave magnetic fields results in large spin-wave intensities which makes efficient 4-magnon-scattering possible at low pumping powers in ultra-thin waveguides. As a result, two or more modes are generated below and above the directly pumped mode with equal frequency spacing in between as demanded by energy conservation. It is shown that microwave power and dc current (SOT) thresholds have to be exceeded before large spin-wave intensities and, therefore, nonlinear modes can be achieved confirming that the onset can be controlled. In order to corroborate the presence of a 4-magnon-scattering process, the appearance of the additional modes are investigated in the time-domain revealing a time delay between directly excited and nonlinear modes. Another indication for a 4-magnon-scattering process was extracted from the spatially resolved spin-wave decay measurements. Due to conservation of momentum during the inelastic scattering one magnon with higher and one with lower momentum compared to the two initial ones has to be generated. At low wave vectors the dispersion relation is flatter resulting in reduced spin-wave propagation. This was indeed observed for the low frequency nonlinear mode. Furthermore, it was shown that nonlinear generation of magnons can be achieved over a wide range of external magnetic fields which might be useful for future applications. The deviation of the frequency behavior of the edge modes was used to exclude them from the explanation of the seen nonlinear effects. The application of this effect might be used in the generation of higher spin-wave frequencies and smaller wave lengths or for the generation of spin-wave frequency combs controlled by pure spin currents. Additionally, the strong increase of spin-wave intensities after the onset of the nonlinear process can be interpreted as a spiking process with clear threshold behavior. This might be used in context of the wide field of neuromorphic computing in which complex and nonlinear device behavior is the key element for classification tasks. 

\subsection*{Acknowledgments}
Financial support by the Deutsche Forschungsgemeinschaft is gratefully acknowledged within program SCHU2922/1-1.  
We thank B. Scheumann for deposition of the Au films. Support by the Nanofabrication Facilities Rossendorf (NanoFaRo) at IBC is gratefully acknowledged.

\subsection*{References}

\bibliographystyle{unsrt}
\bibliography{Paper-data-base}

\begin{thebibliography}{10}

\bibitem{arxiv.2301.06727}
Giovanni Finocchio, Supriyo Bandyopadhyay, Peng Lin, Gang Pan, J.~Joshua Yang,
  Riccardo Tomasello, Christos Panagopoulos, Mario Carpentieri, Vito Puliafito,
  Johan {\AA}kerman, Hiroki Takesue, Amit~Ranjan Trivedi, Saibal Mukhopadhyay,
  Kaushik Roy, Vinod~K. Sangwan, Mark~C. Hersam, Anna Giordano, Huynsoo Yang,
  Julie Grollier, Kerem Camsari, Peter Mcmahon, Supriyo Datta, Jean~Anne
  Incorvia, Joseph Friedman, Sorin Cotofana, Florin Ciubotaru, Andrii Chumak,
  Azad~J. Naeemi, Brajesh~Kumar Kaushik, Yao Zhu, Kang Wang, Belita Koiller,
  Gabriel Aguilar, Guilherme Tempor{\~a}o, Kremena Makasheva, Aida~Tordi
  Sanial, Jennifer Hasler, William Levy, Vwani Roychowdhury, Samiran Ganguly,
  Avik Ghosh, Davi Rodriquez, Satoshi Sunada, Karin Evershor-Sitte, Amit Lal,
  Shubham Jadhav, Massimiliano Di~Ventra, Yuriy Pershin, Kosuke Tatsumura, and
  Hayato Goto.
\newblock Roadmap for unconventional computing with nanotechnology, 2023.

\bibitem{Dieny:2020aa}
B.~Dieny, I.~L. Prejbeanu, K.~Garello, P.~Gambardella, P.~Freitas,
  R.~Lehndorff, W.~Raberg, U.~Ebels, S.~O. Demokritov, J.~Akerman, A.~Deac,
  P.~Pirro, C.~Adelmann, A.~Anane, A.~V. Chumak, A.~Hirohata, S.~Mangin,
  Sergio~O. Valenzuela, M.~Cengiz Onba{\c s}lı, M.~d'Aquino, G.~Prenat,
  G.~Finocchio, L.~Lopez-Diaz, R.~Chantrell, O.~Chubykalo-Fesenko, and
  P.~Bortolotti.
\newblock {Opportunities and challenges for spintronics in the microelectronics
  industry}.
\newblock {\em Nature Electronics}, 3(8):446--459, 2020.

\bibitem{Christensen_2022}
Dennis~V Christensen, Regina Dittmann, Bernabe Linares-Barranco, Abu Sebastian,
  Manuel~Le Gallo, Andrea Redaelli, Stefan Slesazeck, Thomas Mikolajick, Sabina
  Spiga, Stephan Menzel, Ilia Valov, Gianluca Milano, Carlo Ricciardi, Shi-Jun
  Liang, Feng Miao, Mario Lanza, Tyler~J Quill, Scott~T Keene, Alberto Salleo,
  Julie Grollier, Danijela Markovi{\'c}, Alice Mizrahi, Peng Yao, J~Joshua
  Yang, Giacomo Indiveri, John~Paul Strachan, Suman Datta, Elisa Vianello,
  Alexandre Valentian, Johannes Feldmann, Xuan Li, Wolfram H~P Pernice, Harish
  Bhaskaran, Steve Furber, Emre Neftci, Franz Scherr, Wolfgang Maass, Srikanth
  Ramaswamy, Jonathan Tapson, Priyadarshini Panda, Youngeun Kim, Gouhei Tanaka,
  Simon Thorpe, Chiara Bartolozzi, Thomas~A Cleland, Christoph Posch, ShihChii
  Liu, Gabriella Panuccio, Mufti Mahmud, Arnab~Neelim Mazumder, Morteza
  Hosseini, Tinoosh Mohsenin, Elisa Donati, Silvia Tolu, Roberto Galeazzi,
  Martin~Ejsing Christensen, Sune Holm, Daniele Ielmini, and N~Pryds.
\newblock 2022 roadmap on neuromorphic computing and engineering.
\newblock {\em Neuromorphic Computing and Engineering}, 2(2):022501, may 2022.

\bibitem{Grollier:2020aa}
J.~Grollier, D.~Querlioz, K.~Y. Camsari, K.~Everschor-Sitte, S.~Fukami, and
  M.~D. Stiles.
\newblock Neuromorphic spintronics.
\newblock {\em Nature Electronics}, 3(7):360--370, 2020.

\bibitem{Kimel_2022}
Alexey Kimel, Anatoly Zvezdin, Sangeeta Sharma, Samuel Shallcross, Nuno
  de~Sousa, Antonio Garc{\'\i}a-Mart{\'\i}n, Georgeta Salvan, Jaroslav Hamrle,
  Ond{\v r}ej Stejskal, Jeffrey McCord, Silvia Tacchi, Giovanni Carlotti,
  Pietro Gambardella, Gian Salis, Markus M{\"u}nzenberg, Martin Schultze,
  Vasily Temnov, Igor~V Bychkov, Leonid~N Kotov, Nicol{\`o} Maccaferri, Daria
  Ignatyeva, Vladimir Belotelov, Claire Donnelly, Aurelio~Hierro Rodriguez,
  Iwao Matsuda, Thierry Ruchon, Mauro Fanciulli, Maurizio Sacchi, Chunhui~Rita
  Du, Hailong Wang, N~Peter Armitage, Mathias Schubert, Vanya Darakchieva, Bilu
  Liu, Ziyang Huang, Baofu Ding, Andreas Berger, and Paolo Vavassori.
\newblock The 2022 magneto-optics roadmap.
\newblock {\em Journal of Physics D: Applied Physics}, 55(46):463003, sep 2022.

\bibitem{Kiselev:2003aa}
S.~I. Kiselev, J.~C. Sankey, I.~N. Krivorotov, N.~C. Emley, R.~J. Schoelkopf,
  R.~A. Buhrman, and D.~C. Ralph.
\newblock {Microwave oscillations of a nanomagnet driven by a spin-polarized
  current}.
\newblock {\em Nature}, 425(6956):380--383, 2003.

\bibitem{Demidov:2012aa}
Vladislav~E. Demidov, Sergei Urazhdin, Henning Ulrichs, Vasyl Tiberkevich,
  Andrei Slavin, Dietmar Baither, Guido Schmitz, and Sergej~O. Demokritov.
\newblock {Magnetic nano-oscillator driven by pure spin current}.
\newblock {\em Nature Materials}, 11(12):1028--1031, 2012.

\bibitem{ISI:000264630400001}
Andrei Slavin and Vasil Tiberkevich.
\newblock {Nonlinear Auto-Oscillator Theory of Microwave Generation by
  Spin-Polarized Current}.
\newblock {\em IEEE Transactions on Magnetics}, 45(4):1875--1918, 2009.

\bibitem{SLONCZEWSKI1996L1}
J.C. Slonczewski.
\newblock Current-driven excitation of magnetic multilayers.
\newblock {\em Journal of Magnetism and Magnetic Materials}, 159(1):L1 -- L7,
  1996.

\bibitem{Zahedinejad:2020aa}
Mohammad Zahedinejad, Ahmad~A. Awad, Shreyas Muralidhar, Roman Khymyn, Himanshu
  Fulara, Hamid Mazraati, Mykola Dvornik, and Johan {\AA}kerman.
\newblock {Two-dimensional mutually synchronized spin {H}all nano-oscillator
  arrays for neuromorphic computing}.
\newblock {\em Nature Nanotechnology}, 15(1):47--52, 2020.

\bibitem{arxiv.2301.03859}
Akash Kumar, Himanshu Fulara, Roman Khymyn, Mohammad Zahedinejad, Mona
  Rajabali, Xiaotian Zhao, Nilamani Behera, Afshin Houshang, Ahmad~A. Awad, and
  Johan {\AA}kerman.
\newblock Robust mutual synchronization in long spin hall nano-oscillator
  chains, 2023.

\bibitem{PhysRevLett.122.097202}
K.~Schultheiss, R.~Verba, F.~Wehrmann, K.~Wagner, L.~K\"orber, T.~Hula,
  T.~Hache, A.~K\'akay, A.~A. Awad, V.~Tiberkevich, A.~N. Slavin,
  J.~Fassbender, and H.~Schultheiss.
\newblock Excitation of whispering gallery magnons in a magnetic vortex.
\newblock {\em Phys. Rev. Lett.}, 122:097202, Mar 2019.

\bibitem{Synchronization_of_spin_Hall_nano-oscillators_to_external_microwave_signals}
V.~E. Demidov, H.~Ulrichs, S.~V. Gurevich, S.~O. Demokritov, V.~S. Tiberkevich,
  A.~N. Slavin, A.~Zholud, and S.~Urazhdin.
\newblock {Synchronization of spin Hall nano-oscillators to external microwave
  signals}.
\newblock {\em Nature Communications}, 5:3179, 2014.

\bibitem{doi:10.1063/1.5082692}
T.~Hache, T.~Weinhold, K.~Schultheiss, J.~Stigloher, F.~Vilsmeier, C.~Back,
  S.~S. P.~K. Arekapudi, O.~Hellwig, J.~Fassbender, and H.~Schultheiss.
\newblock {Combined frequency and time domain measurements on injection-locked,
  constriction-based spin Hall nano-oscillators}.
\newblock {\em Applied Physics Letters}, 114(10):102403, 2019.

\bibitem{PhysRevApplied.13.054009}
T.~Hache, M.~Va\ifmmode~\check{n}\else \v{n}\fi{}atka,
  L.~Flaj\ifmmode~\check{s}\else \v{s}\fi{}man, T.~Weinhold, T.~Hula,
  O.~Ciubotariu, M.~Albrecht, B.~Arkook, I.~Barsukov, L.~Fallarino, O.~Hellwig,
  J.~Fassbender, M.~Urb\'anek, and H.~Schultheiss.
\newblock {Freestanding Positionable Microwave-Antenna Device for
  Magneto-Optical Spectroscopy Experiments}.
\newblock {\em Phys. Rev. Applied}, 13:054009, May 2020.

\bibitem{Wagner:2018aa}
Kai Wagner, Andrew Smith, Toni Hache, Jen-Ru Chen, Liu Yang, Eric Montoya,
  Katrin Schultheiss, J{\"u}rgen Lindner, J{\"u}rgen Fassbender, Ilya
  Krivorotov, and Helmut Schultheiss.
\newblock Injection locking of multiple auto-oscillation modes in a tapered
  nanowire spin {H}all oscillator.
\newblock {\em Scientific Reports}, 8(1):16040, 2018.

\bibitem{doi:10.1063/1.4871519}
V.~E. Demidov, S.~Urazhdin, A.~B. Rinkevich, G.~Reiss, and S.~O. Demokritov.
\newblock Spin hall controlled magnonic microwaveguides.
\newblock {\em Applied Physics Letters}, 104(15):152402, 2014.

\bibitem{doi:10.1063/5.0090033}
T.~Hula, K.~Schultheiss, F.~J.~T. Gon{\c c}alves, L.~K{\"o}rber, M.~Bejarano,
  M.~Copus, L.~Flacke, L.~Liensberger, A.~Buzdakov, A.~K{\'a}kay, M.~Weiler,
  R.~Camley, J.~Fassbender, and H.~Schultheiss.
\newblock Spin-wave frequency combs.
\newblock {\em Applied Physics Letters}, 121(11):112404, 2022.

\bibitem{https://doi.org/10.48550/arxiv.2208.09036}
Mauricio Bejarano, Francisco J.~T. Goncalves, Toni Hache, Michael Hollenbach,
  Christopher Heins, Tobias Hula, Lukas K{\"o}rber, Jakob Heinze, Yonder
  Berenc{\'e}n, Manfred Helm, J{\"u}rgen Fassbender, Georgy~V. Astakhov, and
  Helmut Schultheiss.
\newblock Nonlinear magnon control of atomic spin defects in scalable quantum
  devices, 2022.

\bibitem{Simon:2022aa}
Brecht~G. Simon, Samer Kurdi, Joris~J. Carmiggelt, Michael Borst, Allard~J.
  Katan, and Toeno van~der Sar.
\newblock Filtering and imaging of frequency-degenerate spin waves using
  nanopositioning of a single-spin sensor.
\newblock {\em Nano Letters}, 22(22):9198--9204, 11 2022.

\bibitem{Iacopo:uz}
Bertelli Iacopo, Carmiggelt~Joris J., Yu~Tao, Simon~Brecht G., Pothoven~Coosje
  C., Bauer Gerrit~E. W., Blanter~Yaroslav M., Aarts Jan, and van~der
  Sar~Toeno.
\newblock Magnetic resonance imaging of spin-wave transport and interference in
  a magnetic insulator.
\newblock {\em Science Advances}, 6(46):eabd3556, 2021/12/02.

\bibitem{Lee-Wong:2020vc}
Eric Lee-Wong, Ruolan Xue, Feiyang Ye, Andreas Kreisel, Toeno van~der Sar, Amir
  Yacoby, and Chunhui~Rita Du.
\newblock Nanoscale detection of magnon excitations with variable wavevectors
  through a quantum spin sensor.
\newblock {\em Nano Letters}, 20(5):3284--3290, 05 2020.

\bibitem{Wang:2020vf}
Xiaoche Wang, Yuxuan Xiao, Chuanpu Liu, Eric Lee-Wong, Nathan~J. McLaughlin,
  Hanfeng Wang, Mingzhong Wu, Hailong Wang, Eric~E. Fullerton, and Chunhui~Rita
  Du.
\newblock Electrical control of coherent spin rotation of a single-spin qubit.
\newblock {\em npj Quantum Information}, 6(1):78, 2020.

\bibitem{Neethirajan:2023aa}
Jeffrey~Neethi Neethirajan, Toni Hache, Domenico Paone, Dinesh Pinto, Andrej
  Denisenko, Rainer St{\"o}hr, P{\'e}ter Udvarhelyi, Anton Pershin, Adam Gali,
  Joerg Wrachtrup, Klaus Kern, and Aparajita Singha.
\newblock Controlled surface modification to revive shallow nv--centers.
\newblock {\em Nano Letters}, 03 2023.

\bibitem{DyakonovPerrel}
M.I. Dyakonov and V.I. Perel.
\newblock {Possibility of Orienting Electron Spins with Current}.
\newblock {\em JETP Letters}, 13(11):467, June 1971.

\bibitem{ISI:000082242600034}
JE~Hirsch.
\newblock {Spin Hall effect}.
\newblock {\em Phys. Rev. Lett.}, 83(9):1834--1837, 1991.

\bibitem{ISI:000324930200001}
Axel Hoffmann.
\newblock {Spin Hall Effects in Metals}.
\newblock {\em IEEE Transactions on Magnetics}, 49(10):5172--5193, 2013.

\bibitem{9427163}
Qiming Shao, Peng Li, Luqiao Liu, Hyunsoo Yang, Shunsuke Fukami, Armin Razavi,
  Hao Wu, Kang Wang, Frank Freimuth, Yuriy Mokrousov, Mark~D. Stiles, Satoru
  Emori, Axel Hoffmann, Johan {\AA}kerman, Kaushik Roy, Jian-Ping Wang, See-Hun
  Yang, Kevin Garello, and Wei Zhang.
\newblock Roadmap of spin--orbit torques.
\newblock {\em IEEE Transactions on Magnetics}, 57(7):1--39, 2021.

\bibitem{doi:10.1063/5.0008988}
T.~Hache, Y.~Li, T.~Weinhold, B.~Scheumann, F.~J.~T. Gon{\c c}alves,
  O.~Hellwig, J.~Fassbender, and H.~Schultheiss.
\newblock {Bipolar spin Hall nano-oscillators}.
\newblock {\em Applied Physics Letters}, 116(19):192405, 2020.

\bibitem{doi:10.1063/5.0054169}
L.~K{\"o}rber, G.~Quasebarth, A.~Otto, and A.~K{\'a}kay.
\newblock Finite-element dynamic-matrix approach for spin-wave dispersions in
  magnonic waveguides with arbitrary cross section.
\newblock {\em AIP Advances}, 11(9):095006, 2021.

\bibitem{TetraX}
Lukas K{\"o}rber, Gwendolyn Quasebarth, Alexander Hempel, Friedrich Zahn,
  Andreas Otto, Elmar Westphal, Riccardo Hertel, and Attila Kakay.
\newblock {TetraX: Finite-Element Micromagnetic-Modeling Package}, January
  2022.

\bibitem{10.3389/fphy.2015.00035}
Thomas Sebastian, Katrin Schultheiss, Bj{\"o}rn Obry, Burkard Hillebrands, and
  Helmut Schultheiss.
\newblock {Micro-focused Brillouin light scattering: imaging spin waves at the
  nanoscale}.
\newblock {\em Frontiers in Physics}, 3:35, 2015.

\bibitem{PhysRevB.102.054422}
Andrew Smith, Kemal Sobotkiewich, Amanatullah Khan, Eric~A. Montoya, Liu Yang,
  Zheng Duan, Tobias Schneider, Kilian Lenz, J\"urgen Lindner, Kyongmo An,
  Xiaoqin Li, and Ilya~N. Krivorotov.
\newblock {Dimensional crossover in spin Hall oscillators}.
\newblock {\em Phys. Rev. B}, 102:054422, Aug 2020.

\end{thebibliography}

\end{document}